\documentclass[onecollarge,natbib]{svjour2}
\bibpunct{[}{]}{;}{n}{}{,} 
\smartqed  
\usepackage{graphicx}

\usepackage[utf8]{inputenc}

\usepackage[T1]{fontenc}

\usepackage{slashed}

\usepackage{times}

\usepackage{epstopdf}

\usepackage{amssymb,amsfonts,amsmath}

\usepackage{dsfont,bbm}

\usepackage{dcolumn}

\usepackage{epsfig}

\usepackage{graphicx}

\usepackage[caption=false]{subfig}

\usepackage{dsfont}

\usepackage{slashed}

\usepackage[active]{srcltx}

\usepackage[usenames]{color}
\newcommand{\up}{{\uparrow}}
\newcommand{\down}{{\downarrow}}

\journalname{Few-Body Systems}
\begin{document}

\title{Light-cone sum rule approach for Baryon form factors
}


\author{Nils Offen
}


\institute{N. Offen \at
              Fakult\"at Physik, Universit\"at Regensburg, D-93040 Regensburg \\
              Tel.: +49-941-9432003
              \email{nils.offen@ur.de}           
}

\date{Received: date / Accepted: date}

\maketitle
\begin{abstract}

We present the state-of-the-art of the light-cone sum rule approach to Baryon form factors.
The essence of this approach is that soft Feynman contributions are calculated in terms
of small transverse distance quantities using dispersion relations and duality.
The form factors are thus expressed in terms of nucleon wave functions at small
transverse separations, called distribution amplitudes, without any additional parameters.
The distribution amplitudes, therefore, can be extracted from the comparison with the
experimental data on form factors and compared to the results of lattice QCD simulations.
\end{abstract}


\section{Introduction}
\label{intro}
Understanding the characteristics of Hadrons in terms of QCD degrees fo freedom, 
namely quarks and gluons, is one of the central challenges of particle physics. 
It is understood that form factors at large momentum transfer $Q^2$ can be described 
in terms of distribution amplitudes, i.e. light-cone wave functions at small light-like 
separation. \cite{Chernyak:1977as,Radyushkin:1977gp,Lepage:1979zb} 
In this way experimental measurements of form factors can be connected to the momentum 
distribution of quarks inside the involved hadrons. For mesons the gold plated modes are 
the so called transition form factors $\pi,\eta^{(\prime)} \to\gamma\gamma*$ where the hard 
formally leading contribution in $\frac{1}{Q^2}$ is not suppressed by powers of $\frac{\alpha_s}{\pi}$. 
But even there power corrections can reach up to $\sim20$ $\%$ at large $Q^2\sim40$ GeV$^2$.\cite{Agaev:2010aq, Mikhailov:2009kf, Bakulev:2011rp}\\
For electromagnetic Baryon form factors the hard contributions are suppressed by 
$\left(\frac{\alpha_s}{\pi}\right)^2\sim0.01$ compared to the so called Feynman (soft) terms where one quark carries 
almost all of the momentum of the parent hadron and interacts solely via soft gluons. Therefore the 
asymptotic regime where the perturbative description in terms of distribution amplitudes is correct 
is postponed to very high $Q^2$ far out of reach of current experiments.\\
Under these circumstances additional model assumptions have to be made to interpret experimental data.
One possibility is to model transverse momentum dependent (TMD) light-cone wave function and use 
Sudakov suppression of large transverse distances as initially suggested by Li and Sterman \cite{Li:1992nu}.\\
The possibility we advocate is called light-cone sum rules. It is based on an light-cone expansion in baryon distribution 
amplitudes of increasing twist using dispersion relations and duality. Soft- and hard contributions are calculated on the 
same footing and there is no double counting \cite{Braun:1999uj}. This method gives up to now the most direct connection 
between form factors and distribution amplitudes and has already been succesfully applied to several meson decays.\\
For baryons the case is more complicated and it shall be discussed in some detail in the next section.

\section{Light-cone sum rules for $N$ and $N^*$ form factors}
\label{sec:1}
\begin{figure}[ht]
\centerline{\includegraphics[width=5cm, clip = true]{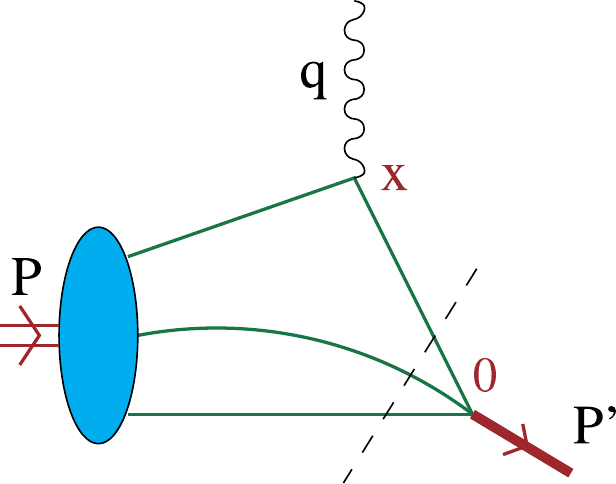}}
\caption{\label{Fig1}
Schematic structure of the light-cone sum rule for baryon form factors.
}
\end{figure}

The basic object of the LCSR approach to baryon form factors~\cite{Braun:2001tj,Braun:2006hz}
is the correlation function
\begin{equation}
\Pi_\mu(P,q)\,=\,\int\! dx\, e^{-iqx}\langle 0| T \{ \eta (0) j_\mu(x) \} | P \rangle
\label{eq:corrfunc}
\end{equation}
where $j_\mu$  represents the electromagnetic probe and $\eta$
is a suitable operator with nucleon quantum numbers.
The other (in this example, initial state) nucleon is explicitly represented by its state vector
$| P\rangle $, see a schematic representation in Fig.~\ref{Fig1}.
LCSRs are then derived by matching two different representations of the correlation function:
If both the momentum flowing through the $\eta$-vertex $P'^2=(P-q)^2$ and the momentum transfer 
$Q^2$ are large and negative it can be shown that the main contribution to the integral in (\ref{eq:corrfunc}) 
comes from the region $x^2\approx0$. Hence it can be studied using the operator product expansion of the time ordered 
product of the two currents around the light-cone. The light-cone divergence of the coefficient function is governed by 
the twist, i.e. dimension minus spin, of the respective operator. The matrix element of the operator is related 
to the baryon distribution amplitude. The resulting expression is then analytically continued to positive $P'^2$ by 
dispersion relations.\\
On the other hand the correlation function can be represented as a complete sum over intermediate hadron states 
and can be written as a dispersion integral in $P'^2$ with the nucleon or $N^*$ contribution separated explicitly from 
the higher states.\\
Quark-Hadron duality allows to equate both representations from a certain duality threshold $s_0$ on giving an expression for the form factor
\[\dfrac{\lambda_1\,F_1(Q^2)}{m_N^2-P'^2}\,=\,\int_0^{s_0}\dfrac{ds}{s-P'^2}\,\mbox{Im}_s \Pi(s,Q^2)\,+\,\mbox{subtractions}.\]
A Borel-transformation gets rid of subtraction terms needed to render the dispersion integrals finite and suppresses higher order states
(contributions of large $s$) on the cost of introducing an additional parameter $M^2$. The dependence on this parameter is artificial 
in a similar way as the dependence on the factorization scale $\mu$ in perturbation theory.

\subsection{Form factors}
\label{sec:1.1}
The electromagnetic or electroproduction form factors we are considering are conventionally defined as 
the matrix element of the electromagnetic current
\begin{equation}
\label{em}
 j_{\mu}^{\rm em}(x) = e_u \bar{u}(x) \gamma_{\mu} u(x) + e_d \bar{d}(x) \gamma_{\mu} d(x)
\end{equation}
taken either between nucleon states or between the negative parity spin $\frac{1}{2}$ partner $N^*$ and a nucleon:
\begin{eqnarray}
\label{F1F2}
\langle N(P')| j_{\mu}^{\rm em}(0)|N(P)\rangle &=&
\bar{u}_N(P')\left[\gamma_{\mu}F_1(Q^2)-i\frac{\sigma_{\mu\nu}q^{\nu}}{2m_N}F_2(Q^2)\right]u_N(P),\nonumber\\
\langle N^*(P') | j_{\mu}^{\rm em} |N(P)\rangle &=&
 \bar{u}_{N^*}(P')\left[\gamma_5 \frac{{G_1(q^2)}}{m_N^2}(\slashed{q} q_\mu - q^2 \gamma_\mu)
 -i \frac{{G_2(q^2)}}{m_N} \sigma_{\mu\rho} q^{\rho}\right]  u_N(P)\,.
\end{eqnarray}
Roughly speaking they are a measure for the probability of a nucleon being hit by an energetic photon to form a nucleon or 
a $N^*$. The second form factor is needed to describe the effect of the anomalous magnetic moment of the respective hadron. Possible third and fourth form factors are not necessary 
due to current conservation and parity invariance of the electromagnetic interaction. For the nucleon these form factors are called 
Dirac $F_1$ and Pauli $F_2$ form factor. For experimental measurements it is more convenient to consider the so called electric and magnetic Sachs form factors 
\begin{equation}
\label{GMGE}
G_M(Q^2) \, = \, F_1(Q^2)+F_2(Q^2),\quad G_E(Q^2) \, = \, F_1(Q^2)-\frac{Q^2}{4m_N^2}F_2(Q^2).
\end{equation}
They lead to a separation of the form factors in the famous Rosenbluth scattering cross-section.\\
The helicity amplitudes
$A_{1/2}(Q^2)$ and $S_{1/2}(Q^2)$ for the electroproduction of $N^*(1535)$
can be expressed in terms of the form factors \cite{Aznauryan:2008us} via:
\begin{equation}
A_{1/2} \,=\, e B
\Big[ Q^2 G_1(Q^2) + m_N(m_{N^*}-m_N)G_2(Q^2) \Big],
\quad
{S}_{1/2}\,=\,  \frac{e BC }{\sqrt{2}}
\Big[(m_{N}\!-\!m_{N^*})G_1(Q^2)+m_{N}G_2(Q^2)\Big],
\label{def:A12S12}
\end{equation}
where $e=\sqrt{4\pi\alpha}$ is the elementary charge and $B$, $C$ are kinematic factors defined as
\begin{equation}
   B \,=\, \sqrt{\frac{Q^2+(m_{N^*}+m_N)^2}{2 m_N^5(m_{N^*}^2-m_N^2)}},\quad
   C \, =\, \sqrt{1+\frac{(Q^2-m_{N^*}^2+m_N^2)^2}{4 Q^2 m_{N^*}^2}} \,.
\end{equation}
\subsection{Distribution Amplitudes}
\label{sec:1.2}
One of the attractive features about light-cone sum rules is that one can calculate form factors in terms 
of distribution amplitudes which correspond to light-cone wave functions at small transverse distances and are fundamental 
process independent functions that describe the longitudinal momentum distribution of the partons inside the hadron.\\
They are defined as matrix elements of non-local light-ray operators. The leading twist distribution amplitude of the nucleon
is given by \cite{Chernyak:1983ej,Braun:2000kw}
\begin{equation}
\label{varphi-N}
\langle 0 |
\epsilon^{ijk}\! \left(u^{\up}_i(a_1 n) C \!\!\not\!{n} u^{\down}_j(a_2 n)\right)
\!\not\!{n} d^{\up}_k(a_3 n)
| P\rangle \,=\, - \frac12 f_N\,Pn\! \not\!{n}\, N^\up(P)\! \!\int\! [dx]
\,e^{-i P n \sum x_i a_i}\,
\varphi_N(x_i)\,,
\end{equation}
where $n$ is a light-like vector $n^2=0$ and $f_N$ is the decay constant of the nucleon. 
The distribution amplitudes can be expanded into a set of orthogonal polynomials which are eigenfunctions of the corresponding one-loop evolution kernel 
\cite{Braun:2003rp, Bukhvostov:1985rn, Braun:2008ia, Braun:2009vc}. The coefficients are matrix elements of local conformal operators which can be calculated on the lattice 
\cite{Braun:2008ur,lattice2013,Braun:2014wpa} to constrain the shape of the respective distribution amplitude, see also table 1 and 2. In the following we will call these coefficients, 
shape parameters.
Higher twist distribution amplitudes either describe Fock-states with additional partons, e.g. $qqqG$-states, or with relative angular momentum or both. \cite{Braun:2000kw} 
For the $N^*$ there is some freedom in defining the distribution amplitudes by choosing different positions of $\gamma_5$. We have defined them in such a way 
that all the relations for the nucleon case stay intact and that the coefficient functions in the light-cone sum rules are exactly the same. Since 
distribution amplitudes with additional partons are up to now very poorly known we don't consider them in our calculation.

\section{Results}
\label{sec:2}
The results presented here needed several prerequisites which were derived in the last several years. 
\begin{enumerate}
\item a consistent and practical renormalization scheme for three-quark operators was developed \cite{Kraenkl:2011qb}
\item expressions for matrix elements of operators with non light-like distance in terms of distribution amplitudes were derived \cite{Anikin:2013aka}
\item next to leading order corrections both for twist 3 and 4 were calculated \cite{Anikin:2013aka}
\item the kinematic contributions to higher twist distribution amplitudes, the so called Wandzura-Wilczek contributions, were taken into account \cite{Anikin:2013yoa}
\item off light-cone corrections ($x^2$-corrections) to leading twist distribution amplitudes were recalculated \cite{Anikin:2013aka}
\item the leading twist distribution amplitude was expanded up to second order \cite{Anikin:2013aka}
\end{enumerate}
These advances allowed for the first time to make quantitative statements on the shape of the nucleon and $N^*$ distribution amplitude based on experimental data. 
The extracted shape parameters for the nucleon and $N^*$ with lattice results for comparison are given in table \ref{table1} and \ref{table2}. 
The shape of the resulting distribution amplitudes is plotted in figure \ref{fig:nucleonDA}.\vspace*{-0.5cm}
\begin{table}[h]

\caption{LCSR (ABO1) and LCSR (ABO2) refer to the two models ABO1 and ABO2 extracted in \cite{Anikin:2013aka}. The values of the normalization of 
the leading and next-to-leading twist distribution amplitudes $f_N/\lambda_1$ and the first order shape parameter of the leading distribution amplitude, $\varphi_{10}$ and $\varphi_{11}$ 
have been fixed before fitting to the experimental data. The comparatively large value of $f_N$ was needed to get the normalization of the experimental data right and is in agreement 
with a recent NLO sum rule determination \cite{Gruber:2010bj}. $\varphi_{20},\,\varphi_{21}$ and $\varphi_{22}$ refer to the second order shape parameters of the leading twist distribution amplitude.
$\eta_{10}$ and $\eta_{11}$ are the first order shape parameter of the twist 4 distribution amplitudes.
All values are given at a scale $\mu^2=2$ GeV$^2$.}
 \begin{center}\scriptsize 
\begin{tabular}{@{}l|l|l|l|l|l|l|l|l|l@{}}\hline
Method &  $f_N/\lambda_1 $ & $\varphi_{10}$ & $\varphi_{11}$ & $\varphi_{20}$ & $\varphi_{21}$ & $\varphi_{22}$ & $\eta_{10}$   & $\eta_{11}$    &  Ref. \\ \hline
LCSR(ABO1) &  $-0.17$         & $0.05$        & $0.05$        & $0.075(15)$   & $-0.027(38)$ & $0.17(15)$    & $-0.039(5)$ & $0.140(16)$   & \cite{Anikin:2013aka} \\ \hline
LCSR(ABO2)  &  $-0.17$         & $0.05$        & $0.05$        & $0.038(15)$   & $-0.018(37)$ & $-0.13(13)$   & $-0.027(5)$ & $0.092(15)$   & \cite{Anikin:2013aka} \\ \hline
LATTICE &  $-0.083(6)$     & $0.043(15)$   & $0.041(14)$   & $0.038(100)$ & $-0.14(15)$  & $-0.47(33)$ & -  & -             & \cite{Braun:2008ur}    \\ \hline
LATTICE &  $-0.075(5)$     & $0.038(3)$    & $0.039(6)$    & $-0.050(80)$  & $-0.19(12)$  & $-0.19(14)$   & -           & -             & \cite{lattice2013}     \\ \hline
     QCDSR (NLO)& $-0.15$         & -             & -             & -             & -            & -             & -           & -             & \cite{Gruber:2010bj}   \\ \hline
\end{tabular}
\end{center}
\label{table1}
\end{table}
\vspace*{-0.5cm}
\begin{table*}[h]
\renewcommand{\arraystretch}{1.2}
\caption[]{ Similar to table \ref{table1}. LCSR (1) corresponds to a fit to the form factors $G_1(Q^2)$ and $G_2(Q^2)$ extracted from the 
measurements of helicity amplitudes in \cite{Aznauryan:2009mx}. The uncertainties of the extracted form factors were added in quadrature.
LCSR (2) is obtained from a fit to helicity amplitudes including all available data at $Q^2>1.7$ GeV$^2$ \cite{Denizli:2007tq,Dalton:2008aa,Armstrong:1998wg,Aznauryan:2009mx}
$\lambda^{N\ast}_1/\lambda^N_1$ and $f_{N^\ast}/\lambda^{N^\ast}_1 $ were fixed to the lattice results.}
\begin{center}\scriptsize
\begin{tabular}{@{}l|l|l|l|l|l|l|l|l|l|l@{}} \hline
Method     & $\lambda^{N\ast}_1/\lambda^{N}_1 $ &$f_{N^\ast}/\lambda^{N^\ast}_1 $ & $\varphi_{10}$ & $\varphi_{11}$ & $\varphi_{20}$ & $\varphi_{21}$ & $\varphi_{22}$ & $\eta_{10}$   & $\eta_{11}$    & Ref. \\ \hline
LCSR (1) & 0.633     &0.027      & 0.36     & -0.95      & 0       & 0        & 0         & 0.00    & 0.94   & \cite{Anikin:2015ita} \\ \hline
LCSR (2) & 0.633     &0.027      & 0.37     & -0.96      & 0       & 0        & 0         & -0.29   & 0.23   & \cite{Anikin:2015ita} \\ \hline
LATTICE  & 0.633(43) &0.027(2)   & 0.28(12)  & -0.86(10)  & 1.7(14) & -2.0(18) & 1.7(26)   & -       & -      & \cite{Braun:2014wpa} \\ \hline
\end{tabular}
\end{center}
\label{table2}
\renewcommand{\arraystretch}{1.0}
\end{table*}
\vspace*{-0.5cm}

\subsection{Nucleon electromagnetic form factors}
\label{sec:2.1}
\begin{figure*}[t]
   \begin{center}
\includegraphics[width= 5.5cm, clip = true]{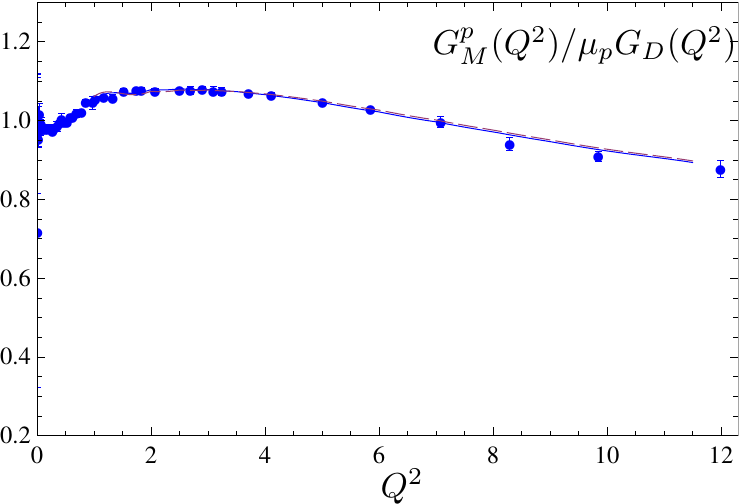}
\qquad
\includegraphics[width= 5.5cm, clip = true]{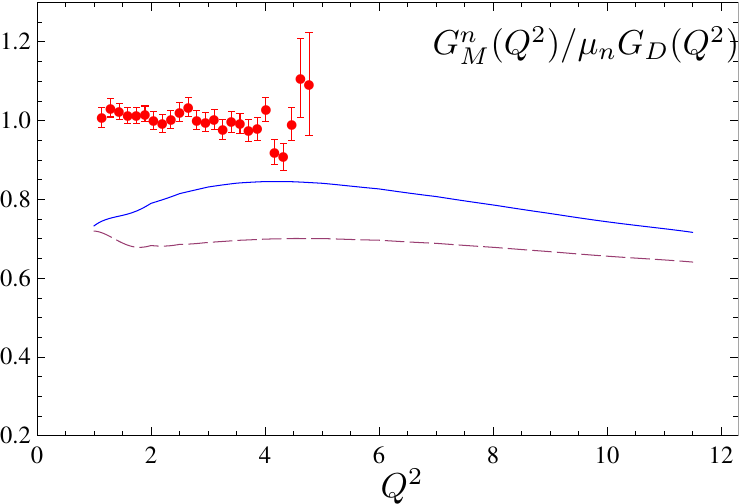}
\\[5mm]
\includegraphics[width= 5.5cm, clip = true]{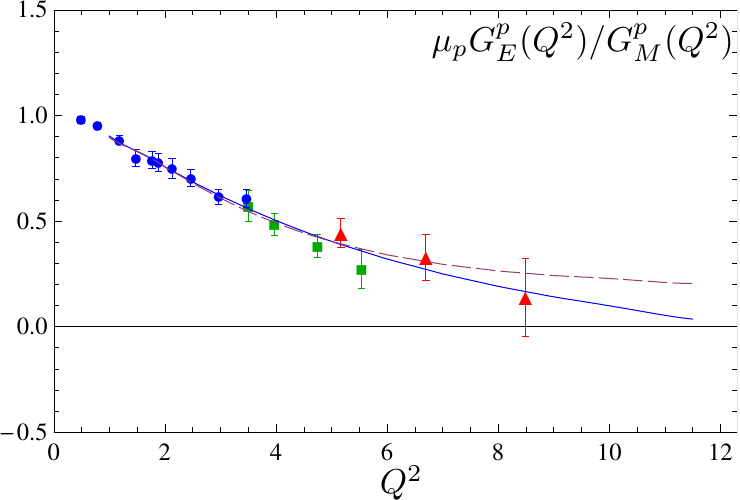}
\hspace*{0.5cm}
\includegraphics[width= 5.5cm, clip = true]{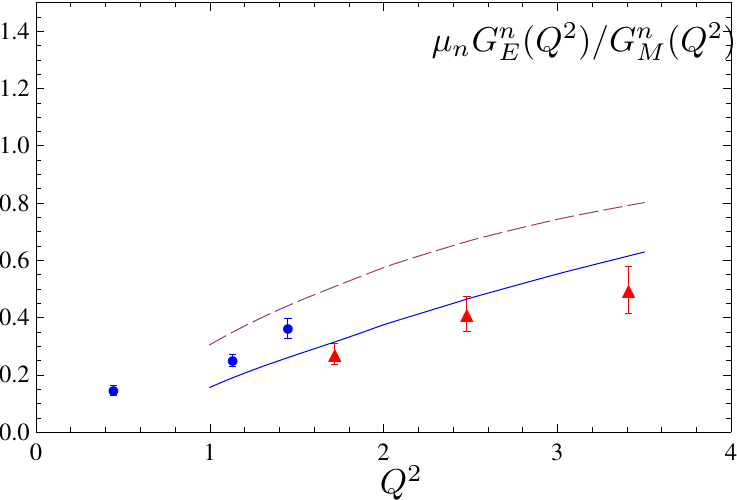}
   \end{center}
\caption{Nucleon electromagnetic form factors from LCSRs compared to the experimental data
\cite{Arrington:2007ux,Lachniet:2008qf,Gayou:2001qd,Punjabi:2005wq,Puckett:2010ac,Plaster:2005cx,Riordan:2010id}.
Parameters of the nucleon DAs correspond to the sets ABO1 and ABO2 in Table~\ref{table1} for the solid and dashed
curves, respectively. The fits were done for different Borel parameters, i.e. $M^2=1.5$~GeV$^2$ for ABO1 and $M^2=2$~GeV$^2$ for ABO2.
}
\label{fig:ABO1}
\end{figure*}

We did two separate fits fixing the normalization $f_N/\lambda_1$, and the lowest order shape parameters $\varphi_{10}$, $\varphi{11}$ to the values given in table \ref{table1} 
for two Borel-prameters to the proton data on the magnetic form factor $G^p_M(Q^2)$ and the ratio 
$G^p_E(Q^2)/G^p_M(Q^2)$ in the interval $1<Q^2<8.5$ GeV$^2$. The fitted values of the shape parameters are given in table 1 and the 
corresponding form factors are shown in figure \ref{fig:ABO1}. Several noteworthy points are seen in the result:\\
First, the experimental data prefers larger values for the ratio of $f_N/\lambda_1$ compatible with NLO sum rule calculations 
but a factor two larger than the lattice result. A fit with all parameters free gets unstable but we see that for different fixed parameters it is a rather 
robust feature that a large normalization and small first order shape parameters are favored.\\
Second, the neutron magnetic form factor $G_M^n(Q^2)$ which is not fitted comes out about 20 to 30 \% too low. This feature is pretty robust. 
A fit to both proton and neutron data simultaneously leads to very large values of $\eta_{10},\,\eta_{11}\sim\mathcal{O}(1)$ and leads to a worse 
description of proton data. We think this is an artefact of missing information on even higher twist distribution amplitudes 
and of the more involved OPE of the form factor $F_2$.
Part of this can be understood with the help of figure \ref{fig:ABO1-DiehlKroll} where the experimental data is 
separated into $u$- and $d$-quark contribution. It is seen that the Dirac form factors $F_1^{u,d}(Q^2)$ are described rather well 
while there are considerable deviations in the Pauli form factors $F_2^{u,d}(Q^2)$ at low $Q^2$. This does not 
come unexpected. At low $Q^2$ one would expect $F_2$ to get sizeable corrections from very high twist, e.g. factorizable 
five quark distribution amplitudes and the structure of the correlation function is so, that to get the same accuracy in the NLO contributions for $F_2$ one 
would need to take into account second order corrections in the deviation from the light-cone which are reserved for a future project. 
Additionally it is seen that the NLO corrections to the $d$-quark contribution are generally very large, probably a feature of the spin-flavor structure of the Ioffe-current,
which means they are generally less precise and potentially stronger affected by higher QCD corrections. Since in the 
neutron the role of the $d$-quark is taken by the $u$-quark, the larger charge factor leads to an enhancement of aforementioned problems 
and therefore to lesser accuracy in the neutron form factors.\\
\begin{figure}[ht]
   \begin{center}
\includegraphics[width= 5.5cm, clip = true]{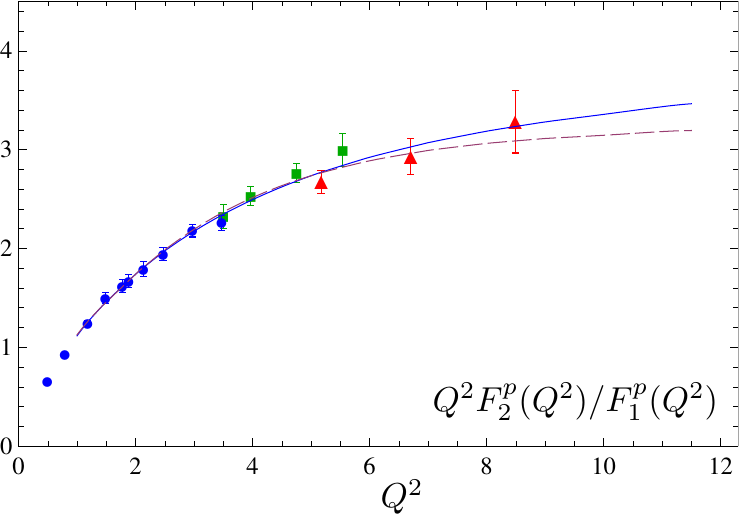}
   \end{center}
\caption{The ratio of Pauli and Dirac electromagnetic proton form factors from LCSRs compared to the experimental data
\cite{Gayou:2001qd,Punjabi:2005wq,Puckett:2010ac}.
Parameters of the nucleon DAs correspond to the sets ABO1 and ABO2 in Table~\ref{table1} for the solid and dashed
curves, respectively. Borel parameter $M^2=1.5$~GeV$^2$ for ABO1 and $M^2=2$~GeV$^2$ for ABO2.}
\label{fig:ABO1-F2F1}
\end{figure}
Third, we did not take into account the uncertainty due to the Borel-parameter separately but rather did two fits with different Borel-parameters. 
The difference in the shape parameters between the two fits can be seen as a measure for the induced deviation. We have illustrated the 
separate variation of the Borel-parameter in figure 6 of \cite{Anikin:2013aka}.
\begin{figure*}[ht]
   \begin{center}
\includegraphics[width= 5.5cm, clip = true]{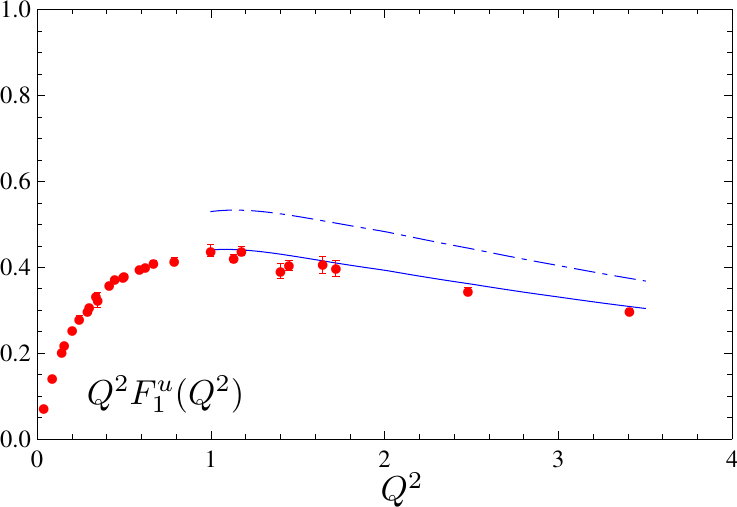}
\qquad
\includegraphics[width= 5.5cm, clip = true]{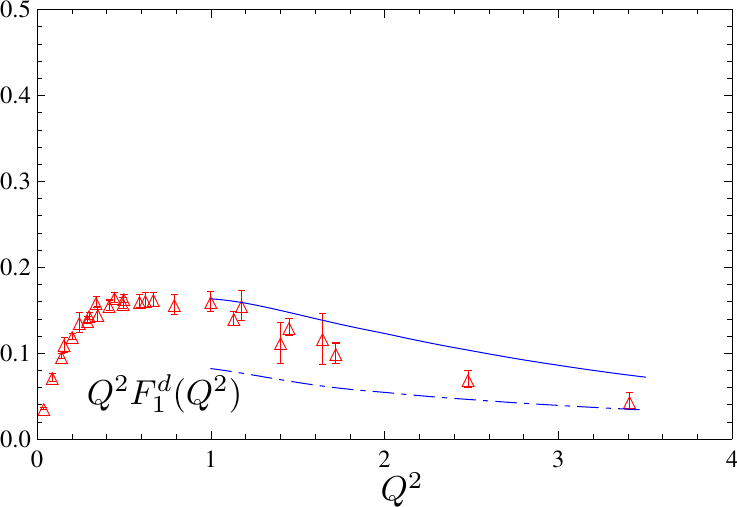}
\\[5mm]
\includegraphics[width= 5.5cm, clip = true]{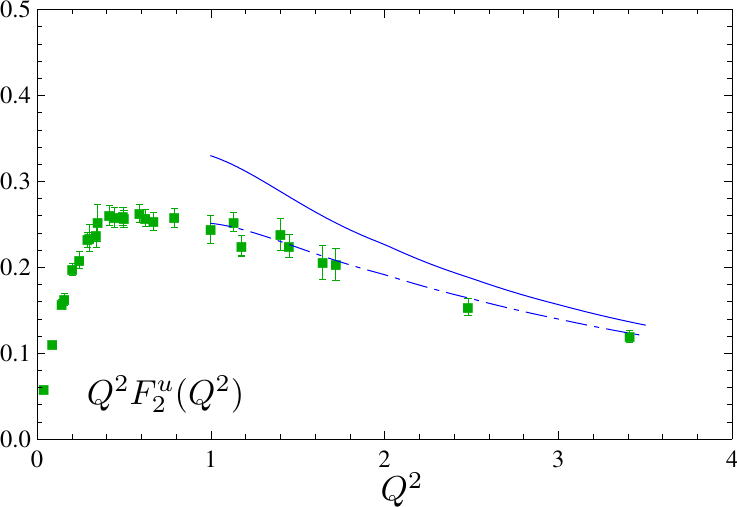}
\hspace*{0.5cm}
\includegraphics[width= 5.5cm, clip = true]{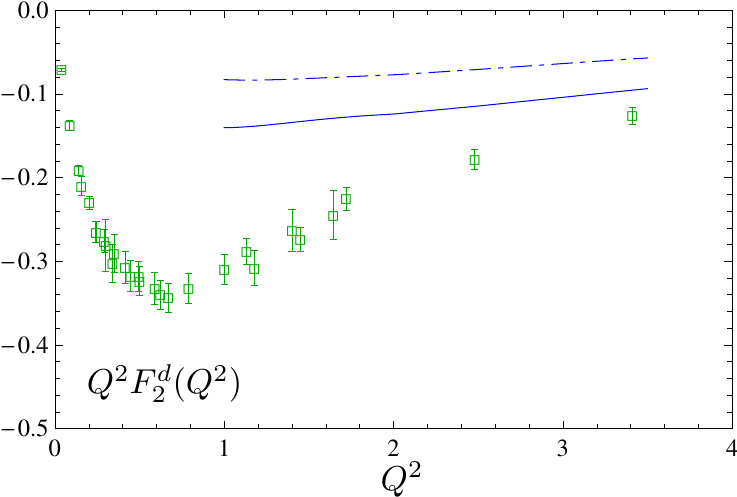}
   \end{center}
\caption{Contributions of different quark flavors to the proton electromagnetic form factors
 compared to the compilation of experimental data in Ref.~\cite{Diehl:2013xca}.
 The corresponding leading-order results are shown by the dash-dotted curves for comparison.
Parameters of the nucleon DAs correspond to the set ABO1 in Table~\ref{table1}.
}
\label{fig:ABO1-DiehlKroll}
\end{figure*}
Fourth, the factorization scale dependence increases with increasing momentum transfer $Q^2$. This might at first glance seem 
counterintuitive but it is consistent with the expected dominant role of the hard scattering corrections which start at 
next-next-to-leading order (NNLO).
\subsection{$N^*$ electroproduction form factors}
\label{sec:2.2}
Due to the larger mass of the $N^*$ the light-cone sum rules get unstable for $Q^2<2$ GeV$^2$ in this case. 
Since data for $Q^2>2$ GeV$^2$ is relatively scarce we set $\varphi_{20},\,\varphi_{21}$ and $\varphi_{22}$ to zero 
and fix $\lambda_1^*$, $f_{N^*}$, $\varphi_{10}$ and $\varphi_{11}$ to the lattice values. In this way we are left only
with the twist 4 parameters $\eta_{10}$ and $\eta_{11}$. We did two separate fits. One where we extracted the form factors 
from the helicity amplitudes $A_{12}$ and $S_{12}$ from \cite{Aznauryan:2009mx} adding the uncertainties in quadrature 
and then fit the shape parameters to the form factors.\\
And one where we fitted directly to all data on the helicity amplitudes. The latter fit is driven by the data from 
\cite{Denizli:2007tq,Dalton:2008aa,Armstrong:1998wg} on the helicity amplitude $A_{12}$ which is not entirely consistent with 
\cite{Aznauryan:2009mx}. Therefore a worse description of the extracted form factors is expected.\\
\begin{figure*}[t]
   \begin{center}
   \includegraphics[width= 5.5cm, clip = true]{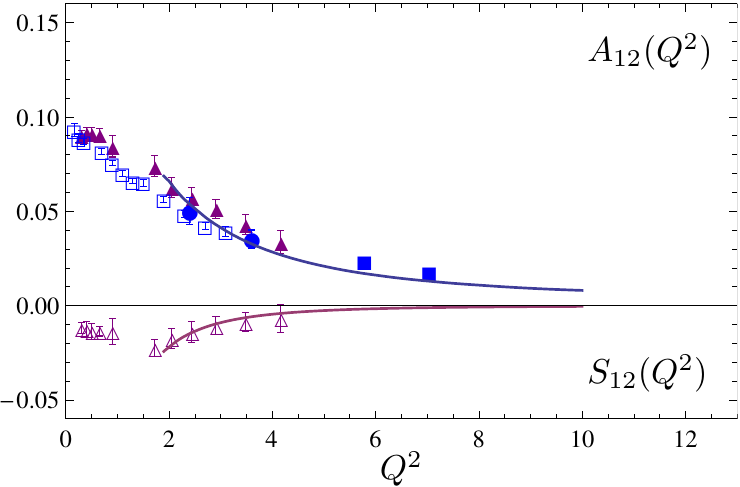}
\qquad
\includegraphics[width= 5.5cm, clip = true]{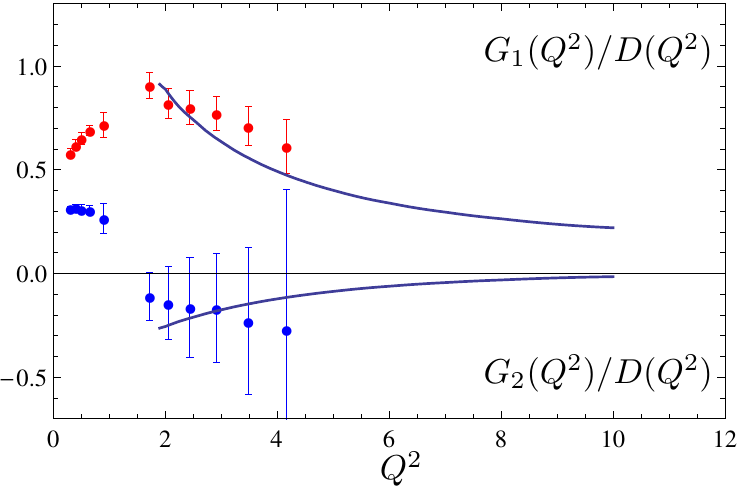}
   \end{center}
\caption{\small\sf Helicity amplitudes $A_{12}$ and $S_{12}$ for electroproduction of $N^*(1535)$ (left panel) and the form factors $G_1(Q^2)$,
$G_2(Q^2)$, normalized to the dipole formula (right panel).
Experimental data on the left panel are taken from \cite{Denizli:2007tq} (empty squares) \cite{Dalton:2008aa} (filled squares)
\cite{Armstrong:1998wg} (filled circles) and \cite{Aznauryan:2009mx} (triangles).
The form factors on the right panel are calculated from the data~\cite{Aznauryan:2009mx} on helicity amplitudes
adding the errors in quadrature.
The curves show the results of the NLO LCSR fit to the form factors $G_1(Q^2)$ and $G_2(Q^2)$ for $Q^2 \ge 1.7~\text{GeV}^2$
with parameters of the $N^\ast(1535)$ DAs specified in the first line in Table 1.
}
\label{fig:ABO1star}
\end{figure*}
In general the sum rules have dominant contributions from P-wave states that is states with one unit of angular momentum. 
Especially the helicity amplitude $A_{12}$ and the Dirac-like form factor $G_1$ are nearly insensitive to the shape of the leading twist distribution amplitude 
mainly due to the very small normalization constant $f_{N^*}$. Even the sensitivity on $\eta_{10}$ and $\eta_{11}$ is rather mild. 
They are predominantly affected by the ratio $\lambda_1^{N^*}/\lambda_1^N$ which comes out rather robust in the range of the lattice result.
$S_{12}$ and the Pauli-like form factor $G_2$ on the other hand are far more sensitive to $\eta_{10}$ and $\eta_{11}$ and due to cancellations of higher 
twist contributions even to the leading twist distribution amplitude but to a lesser degree. More data will be needed to make this extraction more robust.
\begin{figure*}[ht]
   \begin{center}
   \includegraphics[width= 5.5cm, clip = true]{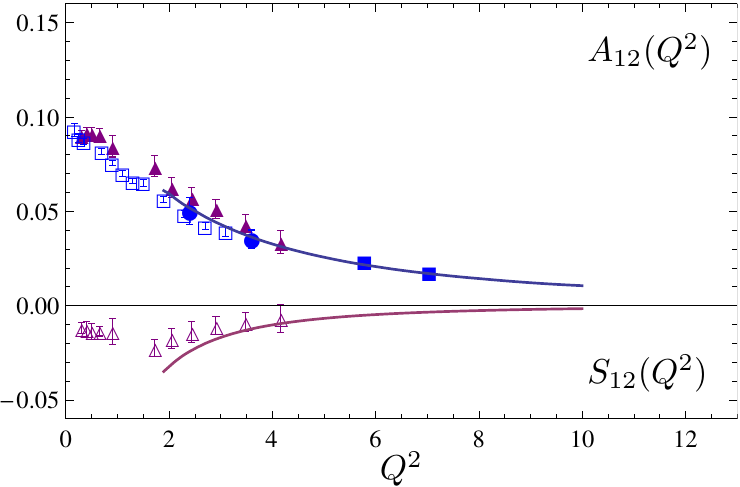}
\qquad
\includegraphics[width= 5.5cm, clip = true]{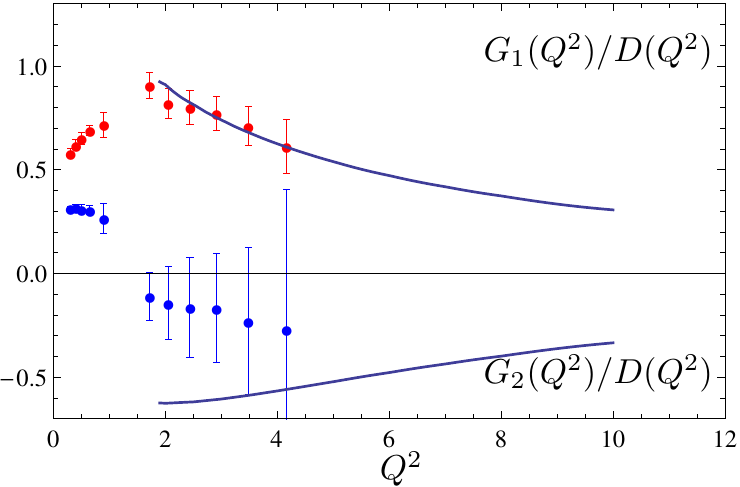}
   \end{center}
\caption{\small\sf The same as in Fig.~\ref{fig:ABO1} but for the fit to helicity amplitudes $A_{12}$, $S_{12}$ including all available data at $Q^2 \ge 1.7~\text{GeV}^2$.
The fitted parameters of the $N^\ast(1535)$ DAs are specified in the second line in Table~1.
}
\label{fig:ABO2star}
\end{figure*}
\section{Conclusions}
\label{sec:3}
We have presented the results of the first consistent NLO light-cone sum rules description of the 
nucleon electromagnetic form factor and of the $N^*$ electroproduction form factors. The results are consistent 
with lattice calculations and are the first quantitative extraction of the leading distribution amplitudes 
from experimental data. For the proton(neutron) a consistent picture emerges, where the distribution amplitude 
peaks for 40\% of the momentum carried by the u(d)-quark with the same helicity as the nucleon and 30\% carried by each of the other quarks.
\begin{figure*}[ht]
\centerline{\includegraphics[width=11cm, clip = true]{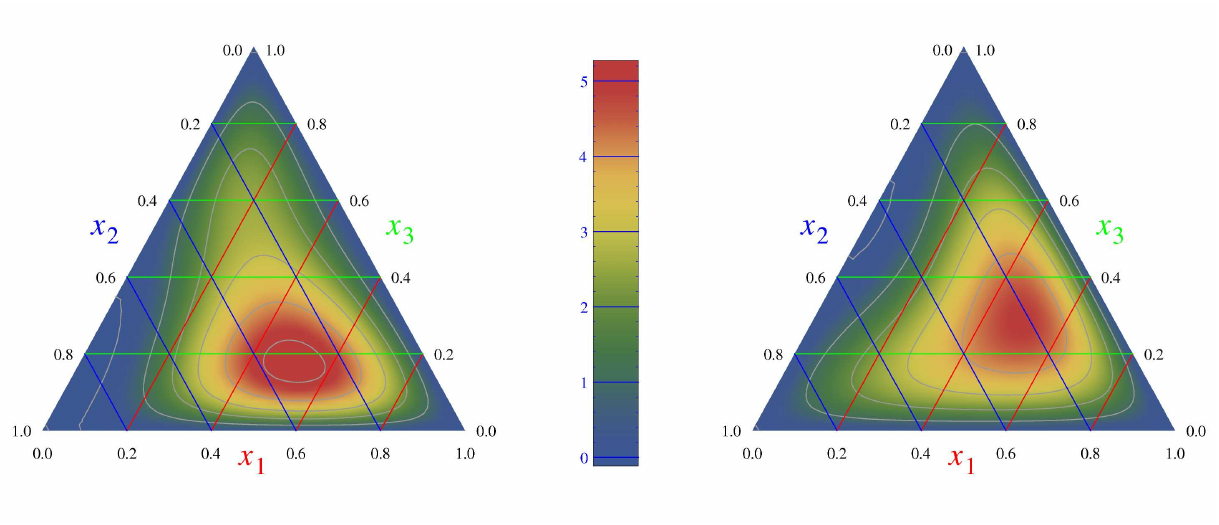}}
\caption{Leading twist distribution amplitude of the proton $\varphi(x_i)$
for the parameter sets ABO1 (left) and ABO2 (right) in Table~\ref{table1}.
Central values are used for the second order parameters.
}
\label{fig:nucleonDA}
\end{figure*}
This is the first hint for a diquark-symmetry coming from a QCD calculation though this symmetry is not exact: It is broken by the different 
renormalization scale behavior of $\varphi_{10}$ and $\varphi_{11}$ and by contributions coming from higher order terms in the conformal expansion.\\
For the $N^*$ the data are described reasonably well, especially for $G_1(Q^2)$, but there are three main problems that increase the uncertainty:
\begin{enumerate}
 \item the small value of $f_{N^*}$ suppresses the leading Fock-state without relative angular momentum and we see that 
 the form factors are dominated by P-wave contributions
 \item the higher mass of the $N^*$ has a similar effect. It increases the contributions of higher twist and it heightens the 
 uncertainty in the NLO part since there we only took into account terms linear in the $N^*$ mass
 \item strong cancellations of higher twist contributions for the Pauli-like form factor $G_2(Q^2)$ 
 \end{enumerate}
Several more projects are either planned or work in progress. The axial form factor of the nucleon and an exploratory study of 
the $\Lambda_{c(b)}\to N^*$ \cite{Emmerich:2016jjm} form factors are close to being finished. An extension towards the Roper-resonance or the $N^*$(1650) 
is planned. In both cases a better understanding of higher twist distribution amplitudes will be needed.\\
Finally on a longer time scale to bring both the form factors $F_2$ and $G_2$ on the same level as $F_1$ and $G_1$ and to lessen the uncertainty 
for the higher masse resonances we plan to calculate the $m_{N(N^*)}^2$ -corrections at NLO. This will require a dedicated calculation where several 
new relations at the twist 5 level will have to be derived.

\end{document}